\documentclass[a4paper]{article}
\usepackage[english]{babel}
\usepackage[dvips]{graphicx}
\usepackage{mathtext}

\usepackage{geometry}
\begin{document}

\title{WMAP2006: Cosmological Parameters and Large-scale Structure of
  the Universe} \author{S.~Apunevych, B.~Venhlovska, Yu.~Kulinich,
  B.~Novosyadlyj}
\maketitle

\medskip

\centerline{\it Astronomical observatory of Ivan Franko National
  University of Lviv} \centerline{\it Kyryla i Methodia, 8, 79005 Lviv
  Ukraine} \medskip

\begin{abstract}
  The parameters of cosmological model with cold dark matter and
  cosmological constant ($\Lambda$CDM) have been determined on a basis
  of three-year cosmic microwave background observations by space
  mission WMAP, as well as the data on the large-scale structure of
  the Universe. The data cover scales from $1$ up to $10000$\,Mpc.
  The best-fit values of $\Lambda$CDM model parameters were found by
  minimization of $\chi^2$ using the Levenberg-Markquardt approach
  ($\Omega_{\Lambda}=0.736\pm 0.065$, $\Omega_m=0.238\pm0.080$,
  $\Omega_{b}=0.05\pm 0.011$, $h=0.68\pm 0.09$, $\sigma_8=0.73\pm
  0.08$ and $n_s=0.96\pm0.015$).  It is shown that the $\Lambda$CDM
  model with these values of the parameters agrees well with the
  angular power spectrum of cosmic microwave background and with power
  spectra of the density perturbations, estimated from spatial
  distributions of galaxies, rich galaxy clusters and from statistics
  of Ly$_{\alpha}$ absorption lines in spectra of distant quasars as
  well. The accordance of modeled characteristics of the large-scale
  structure with observable ones was analyzed, and possible reasons of
  significant discrepancies between some of them were considered.

  \textbf{Keywords:} {CMB temperature fluctuations, cosmological
    parameters, large-scale structure of Universe}

\end{abstract}

\section*{Introduction}

In 2001, the spacecraft of WMAP (Wilkinson Microwave Anisotropy Probe)
had been set into circumsolar orbit at Lagrange point $L_{2}$. It
started full-sky measurements of temperature fluctuations of the
cosmic microwave background (CMB) with resolution of $\approx 13'$. In
2003 the results were published of processing data accumulated during
the first year of observations
\cite{bennett2003,hinshaw2003,spergel2003,verde2003}. This event
signalized the beginning of a new stage in cosmology, the epoch of the
precision cosmology. It was for the first time when the power spectrum
of the CMB temperature fluctuations was determined in the widest range
of angular scales from 20$'$ to 180$^o$, this is equivalent to the
interval of spherical harmonics $2\le \ell \le 1000$. The WMAP data
have provided us with credible confirmation of acoustic peaks
existence and possibility to determine their positions and amplitudes.
The acoustic peaks had been discovered at first in the balloon-borne
experiments like BOOMERANG \cite{deBernardis2000,netterfield02},
MAXIMA \cite{hanany2000,lee2001} and by ground-based interferometer
DASI \cite{halverson2002}. This ``peaked'' structure of the spectrum
was an ultimate argument for the adiabatic character of primordial
perturbations of matter density and space-time metrics. From these
perturbations the galaxies and large-scale structure of the Universe
have been formed. The relations between acoustic peaks amplitudes and
positions within angular power spectrum indicate that $\Lambda$CDM
model with the scale-invariant spectrum of the primordial scalar
density perturbations properly describes the observed Universe.

The bounds were set for the values of six main parameters of the model
\cite{spergel2003,novos2005b} on the basis of the WMAP data
complemented by data set on the large-scale structure of the Universe,
expansion rate and dynamics, abundance of light elements:
$\Omega_{\Lambda}=0.7$--$0.8$, $\Omega_m=0.23$--$0.31$,
$\Omega_{b}=0.04$--$0.05$, $h=0.68$--$0.75$, $A_s=0.75$--$0.92$,
$n_s=0.9$--$0.96$ (the exact values depend on the set of the
observational data used). From these determinations the space
curvature follows to be close to zero, i.e. $\Omega_{k}=0$--$0.04$.
In addition to the main parameters the constraint on the optical depth
to the last scattering surface $\tau$ also was found at a level of
$\tau = 0.23$.  The optical depth is caused by reionization of
intergalactic environment by the first stars. Also the upper bounds
were determined for allowable values of two other parameters, namely
the amplitude of tensor mode of perturbations ($A_t$) and the mass
density of neutrino in units of the critical ($\Omega_{\nu}$), see
\cite{novos2005b,apus2005,novos2005a,spergel2003}.

On March, 2006 the WMAP team had released the data of three-year
observations of CMB temperature fluctuations
\cite{jarosik2006,hinshaw2006,page2006,spergel2006}. These data differ
from previous due to advanced techniques used for treatment of the
measurements noise and foreground contamination. As a result the
signal-to-noise ratio was greatly improved so that the errors for each
point of the CMB power spectrum have decreased as much as twice and
correspondingly the errors of the determination of the amplitudes and
positions of acoustic peaks and dips (troughs) have decreased two
times too.  The range of the allowable values considerably narrowed
for the $\Lambda$CDM model parameters determined on basis of this
spectrum \cite{spergel2006}.  The anisotropy of the CMB polarization
was measured. This polarization was caused by rescattering of the CMB
photons by free electrons generated by ionization of protogalaxies
medium with the first generation of stars. The WMAP have detected the
signal of $E$-mode polarization at lower spherical harmonics $\ell
\approx 2\ldots6$, $\ell(\ell+1)C^{EE}_\ell=0.086\pm 0.029 (\mu K)^2$.
It means the complete reionization took place at $z\approx 8$--$12$ and
the optical depth to the last scattering surface is $\tau=0.09\pm0.03$
\cite{page2006}.

In this paper we determine the main parameters of the $\Lambda$CDM
model using the new data on the amplitudes and positions of acoustic
peaks and characteristics the large-scale structure of the Universe.
Also we shall investigate the concordance between various
observational data and models.

\section{Power spectrum of CMB temperature fluctuations}

\begin{figure}
\begin{center}
  \includegraphics[width=12cm]{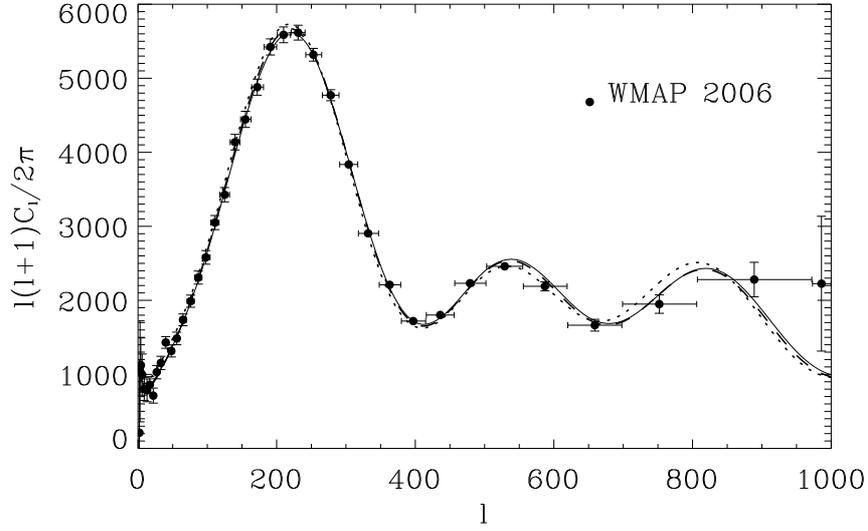}
  \caption{Angular power spectrum of CMB temperature fluctuations
    evaluated according the WMAP experiment data of three years of
    continuous observations.  Solid line denotes the spectrum in
    best-fit $\Lambda$CDM model according to the data of
    \cite{spergel2006} ($\chi^2=37.8$). The dotted line denotes the
    spectrum for $\Lambda$CDM model which is the best-fit to 1-year
    WMAP data and large-scale structure observables
    \cite{novos2005b,apus2005} ($\chi^2=68.2$). The dashed line
    depicts the spectrum of $\Lambda$CDM3 model with parameters found
    further in this paper ($\chi^2=37.2$).}
  \label{wmap3-sp}
\end{center}
\end{figure}

The angular power spectrum evaluated by processing the three-year
observations of the CMB is plotted in Fig.~\ref{wmap3-sp}\footnote{the
  observation data, tables and maps, are available at
  http://lambda.gsfc.nasa.gov/}. The whole spherical harmonics range
$\ell=2$--$1000$ is binned into 39 bins (marked by horizontal bars),
the value of the amplitude was determined for each bin by special
techniques with use of the full-sky map. The sources of foreground
contamination were eliminated during this processing, as well as
traces of Galaxy plane (see \cite{hinshaw2006, jarosik2006} for
details on data processing).  The errors include measurements noise,
guiding errors, signal calibration errors and statistic error
connected to unremovable sample incompleteness (cosmic variance).  The
solid line designates the power spectrum computed with use of CMBFast
code \cite{cmbfast99} for $\Lambda$CDM model with parameters giving
the best-fit to the observed spectrum, $\Omega_{\Lambda}=0.76$,
$\Omega_m=0.24$, $\Omega_{b}=0.042$, $h=0.73$, $A_s=0.83$,
$n_s=0.958$. The goodness of fit can be estimated by statistics of
$\chi^2$, the sum of squared differences between observed and modelled
amplitudes of the spectrum divided by the value of variance for a
central point of each bin.  Assuming no correlation between the values
of power in adjacent bins we get $\chi^2=37.8$ for 33 degrees of
freedom. Also for comparison we have plotted by dotted line the
spectrum of $\Lambda$CDM model with values of parameters given as
best-fit to the data of first-year WMAP observations and large-scale
structure of the Universe \cite{novos2005b,apus2005}, $\chi^2=68.2$
for this spectrum.

The positions and amplitudes of acoustic peaks and dips are important
features of the angular power spectrum. As one can see, using the data
of 2006 we can reliably determine the positions and amplitudes of first
and second peaks and dip between them. The position of the third peak
can not be determined reliably from these data, since resolution of
the WMAP telescope is $\approx 13'$. Only involving the data of
ground-based interferometric and baloon-borne experiments with better
angular resolution makes possible to determine the position and
amplitude of third acoustic peak. However, WMAP2006 data alone 
indicate on power increase beyond the second peak.  The numerical
values for the peaks and dips positions and amplitudes are listed in
Table~\ref{tab_peaks} based on three-year WMAP observations
\cite{hinshaw2006}. Also their values from first-year data release are
quoted therein for comparison \cite{bennett2003}.

\begin{table}
  \caption{The positions and amplitudes of acoustic peaks and dips 
    in the CMB power spectrum as estimated by first-year and 
    three-year WMAP observations.}
  \begin{center}
    \begin{tabular}{|c|c|c|c|c|}
      \hline 
      & & WMAP2003 & & WMAP2006 \\
      \hline
      & & & &\\
      Name&Position&Amplitude&Position&Amplitude \\
      &$l$&$(\Delta T)^2$ $(\mu K)^2$&$l$&$(\Delta T)^2$ $(\mu K)^2$ \\
      & & & &\\
      \hline 
      & & & &\\
      1st peak&$220.1\pm0.8$ &$5583\pm73$ &$220.7\pm0.7$&$5619\pm30$ \\
      1st deep&$411.0\pm3.5$ &$1681\pm41$ &$412.8\pm1.9$&$1704\pm27$ \\
      2nd peak&$546.0\pm10.0$ &$2381\pm83$ &$531.3\pm3.5$&$2476\pm40$ \\
      2nd deep&$ $ & &$674.6\pm12.1$&$1668\pm85$ \\
      3rd peak&$820.0\pm18.0^{ *)}$ &$2150\pm860^{ *)}$ &$1143\pm167^{ **)}$&$2442\pm355^{ **)}$ \\
      & & & &\\
      \hline 
    \end{tabular}
  \end{center}
  $^{ *)}$ +BOOMERANG+ MAXIMA+DASI,  $^{ **)}$ +CBI +ACBAR
  \label{tab_peaks}
\end{table}

As one can see, the precision of amplitudes determination has grown
almost two times. For the first (primary) peak it is better than
$1\%$, for the other elements (except for the third peak) it is better
than $2\%$. In order to estimate the position and amplitude of the
third peak the observations data from CBI \cite{cbi} and
ACBAR\cite{acbar} experiments have been used. The CMB temperature
fluctuations were measured in these experiments with high angular
resolution over the separated areas on the sky. However, it could not
be excluded that given precision of the third peak position and
amplitude determination at the level of $\approx 15\%$ does not take
into account properly the systematic errors, as it follows from
comparison the curve in Fig.~\ref{wmap3-sp} with data on the peak
position from Table.~\ref{tab_peaks}.

The Table~\ref{tab_lowl} lists the values of $(\Delta
T)^2_\ell=\ell(\ell+1)C_\ell/2\pi$ for the range of spherical
harmonics $2\le \ell\le 10$ (this segment belongs to the Sachs-Wolfe
'plateau'), as resulting from data of one-year and three-year
observations. At these scales the main contribution to the uncertainty
of amplitudes of real spectrum comes from cosmic variance, which is
unavoidable because of incomplete statistics of the large areas on the
sky (at spherical harmonics $\ell\le 10$).  However, this part of the
spectrum is particularly valuable for analysis since the information
on the primordial space-time metrics perturbations is available
precisely therein, and it is not distorted by any later effects of the
spectrum shape changes.

We use these data to determine the main parameters of the $\Lambda$CDM
model complementing them by the data on amplitudes of the spectrum at
low harmonics $\ell=2 \ldots 10$, as listed in Table 7 in the paper
\cite{hinshaw2006}.

\begin{table}
  \caption{Amplitudes of the CMB power spectrum at low harmonics 
    according to the data of 1st and 3-years of the WMAP observations.}
  \begin{center}
    \begin{tabular}{|c|c|c|}
      \hline 
      & WMAP2003 & WMAP2006 \\
      \hline
      & & \\
      $\ell$&$(\Delta T)^2_\ell$ $(\mu K)^2$&$(\Delta T)^2_\ell$ $(\mu K)^2$ \\
      \hline 
      & & \\
      2&$123\pm 763$&$211\pm 860$\\
      3&$612\pm 608$&$1041\pm 664$\\
      4&$757\pm 504$&$731\pm 537$\\
      5&$1257\pm 432$&$1521\pm 453$\\
      6&$696\pm 380$&$661\pm 395$\\
      7&$830\pm 342$&$ 1331\pm 353$\\
      8&$628\pm 314$&$671\pm 322$\\
      9&$815\pm 292$&$631\pm 298$\\
      10&$618\pm 276$&$751\pm 280$\\
      & & \\
      \hline 
    \end{tabular}
  \end{center}
  \vspace{0.3cm}
  \label{tab_lowl}
\end{table}

\section{WMAP2006: Cosmological parameters}

The analysis of full set of observations yields indications that the
simplest variant of cosmological model capable to reconcile a
heterogeneous information is the $\Lambda$CDM model. It has main
parameters, such as the Hubble parameter $H_0$ (or its dimensionless
counterpart $h\equiv H_0$/100km/s/Mpc), the value of the cosmological
constant $\Lambda$ (conventionally presented in the units of critical
density, $\Omega_{\Lambda}=\Lambda/3H_0^2$), baryon matter density in
units of critical density $\Omega_b$, matter density (baryon+cold dark
matter) $\Omega_{m}$, the amplitude of primordial spectrum of scalar
perturbations and its tilt $n_s$. The primordial spectrum $P_s(k)=A_s
k^{n_s}$ is Fourier transformation of the two-point spatial
correlation function of the matter fluctuations.  Quite commonly
instead of the power spectrum amplitude $A_{s}$ the more illustrative
quantity $\sigma_8$ is used, the r.m.s amplitude of matter
perturbations at the rich cluster scale $R=8h^{-1}$Mpc,
$$
\sigma_8^2=A_s\int_0^\infty T(k)^2k^{n_s+2}W^2(8k)dk/(2\pi)^{3/2},
$$ 
where $T(k)$ is transfer function dependent on main parameters of the
model, $W(8k)$ is Fourier transformation of the window function
selecting out the volume with the mass of the rich galaxy cluster.

Another important parameter is an optical depth to the last scattering
surface $\tau$ originated by the secondary reionization of baryon
matter by the first luminous objects, massive stars and quasars.  The
$\Lambda$CDM model can be extended to include the tensor mode of
perturbations with the primordial spectrum $P_t(k)=A_t k^{n_t}$ and
contribution of hot dark matter consisting of $N_{\nu}$ species of
neutrinos with non-zero rest mass, $\Omega_{\nu}=\Sigma
m_{\nu}/93.104h^2$. Thus, the total number of independent parameters
of cosmological model can reach 11. Nevertheless not all of them are
really independent, for example the parameter of space curvature is
determined by the matter content and the cosmological constant value:
$\Omega_k=1-\Omega_m-\Omega_{\Lambda}$. The upper constraints for
$A_t$ and $\Omega_{\nu}$ have been determined in a number of recent
papers on determination of the cosmological parameters. The best-fit
values for them appeared so close to zero that we can neglect these
parameters when predicting the properties of large-scale structure of
the Universe. So that we concentrate efforts on determination of main
parameters of the $\Lambda$CDM model, namely the $\Omega_{\Lambda}$,
$\Omega_{m}$, $\Omega_b$, $h$, $A_s$, $n_s$.  We have fixed the value
of the optical depth at $\tau=0.09$ according to the results of the
determination based on the CMB polarization fluctuations detected in
the WMAP experiment \cite{page2006}.

The determination of the $\Lambda$CDM parameters proceeds in the
following steps. Let us have $N$ measured values of the
characteristics of the large-scale structure, and the values of $n$
cosmological parameters must be found on this basis. We assume the
probability distribution function for perturbations to be Gaussian and
consider all observational data as mutually independent.  The number
of degrees of freedom for such system is $\nu=N-n$. The cosmological
parameters are found by means of non-linear Levenberg-Markquardt
minimization applied to the value of
\begin{equation}
  \chi^{2}=\sum_{j=1}^{N}\left({\tilde y_j-y_j \over \Delta \tilde y_j}
  \right)^2~,
\end{equation}
here $\tilde y_j$ is measured value of some $j$-th characteristics of
large-scale structure, $y_j$ is corresponding theoretical prediction,
$\Delta \tilde y_j$ is statistical error of the measured value, and
$N$ is a total number of the observational characteristics.  The
accuracy of the model parameters determination depends not only on the
precision of the measurements but on the accuracy of theoretical
predictions $y_j$ calculation as well. One can use the code CMBFast
\cite{cmbfast96,cmbfast99} or its modifications CAMBCODE
\cite{cambcode}, CMBEasy \cite{cmbeasy} to compute the power spectrum
of the CMB temperature fluctuations and the power spectrum of the
matter density perturbations. The system of the linear
Einstein-Boltzmann equations for perturbations in multi-component
medium is numerically solved in these codes.  They provide inner
accuracy up to $1\%$ or better when evaluating CMB temperature
fluctuations and the matter power spectra. The time required for the
code to compute the single model is quite short, however its direct
application in optimization problems requires significant
computational resources since the large number of the models need to
be calculated during minimum search. As alternative the
semi-analytical methods or interpolations over the beforehand
calculated grid can be used to evaluate the corresponding
predictions. Precisely these methods are used in this paper. The
accuracy is controlled by CMBfast code.

We evaluated peaks positions and amplitudes for predefined set of
cosmological parameters in the following way. First, we complemented
the CMBFast code by subroutine of searching extrema within domain the
CMB power spectrum containing acoustic peaks. The extrema are output
to file along with the corresponding values of parameters
$\Omega_{\Lambda}$, $\Omega_m$, $h$, $\Omega_{b}$, $n_s$.  At the next
stage we construct the grid of values for amplitudes and positions of
acoustic peaks and dips covering the region of parametric space
$\Omega_{\Lambda}=0.0$--$0.8$, $\Omega_m=0.2$--$0.8$, $h=0.3$--$0.9$
(with step equal 0.1) and $\Omega_{b}=0.02$--$0.08$, $n_s=0.9$--$1.1$
(with step equal 0.01), for three values of curvature
$\Omega_k=-0.05;0.0;0.05$.  In every grid node the $A_s$ was
determined by normalization of $\Delta T/T$ power spectrum according
to \cite{bunn97}, with later renormalization within optimization
algorithm. Note, that the renormalization coefficient never get out of
the $0.9$--$1.1$ range. The acoustic peaks/dips positions and
amplitudes for the values of parameters lying between the grid nodes
were found by interpolation using 5-dimensional surface of second
order. The comparison of directly calculated by CMBFast results has
shown that the deviations between interpolated and precise values did
not exceed 0.5\%.

The amplitude of the CMB temperature fluctuations power spectrum at
low spherical harmonics was calculated using the semi-analytical
techniques from \cite{dna2003}, the accuracy was quite satisfactory
for this region.

Besides the WMAP data we also used the data form other cosmological
observations in cosmological parameters determination, namely the
constraints on the Hubble constant $h=0.72\pm0.08$
\cite{freedman2001}, on baryon content $\Omega_bh^2=0.0214\pm0.002$
\cite{kirkman2003} and content of dark matter
$\Omega_m-0.75\Omega_{\Lambda}=-0.25\pm 0.125$ \cite{per99} (these
datasets are denoted ``h'', ``BBN'' and ``SNIa'' accordingly). Also
data on of spatial distribution of galaxies and clusters, peculiar
velocities, mass and X-ray temperature function of rich galaxy
clusters, $Ly_{\alpha}$-clouds in the intergalactic space were
included ((LSS dataset). The list of observables consists of 112
experimental quantities with $1\sigma$ errors (see Tabl.~\ref{tab_lss}
and Fig.~\ref{cl_ly}--\ref{sdss}). We consider all measurements as
statistically independent also assume that the probability
distribution function of experimental errors obeys to the normal law.
The detailed description of the procedure used for calculating the
predictions of large-scale structure characteristics for the given
initial power spectrum of density perturbations is given in our papers
\cite{dn2001,dna2003}. The transfer function $T(k)$ of the initial
power spectrum of density perturbations in $\Lambda$CDM model was
computed with analytical approximation from \cite{eh1999}.

\begin{table}[ht]
  \caption{The best-fit values of parameters of $\Lambda$CDM model according 
    to different determinations}
  \begin{center}
    \begin{tabular}{|c|c|c|c|c|}
      \hline 
      Parameters&Range of the best-fit &WMAP2006 &WMAP2006+&WMAP2006+\\
      &values from \cite{spergel2006}& &BBN+h+SNIa&BBN+h+SNIa+LSS\\
      & &$\Lambda$CDM1 &$\Lambda$CDM2 &$\Lambda$CDM3\\
      \hline
      & & & &\\
      $\Omega_k$&-(0.003-0.04)&-0.003&-0.004& -0.014  \\
      $\Omega_{\Lambda}$&0.65-0.76& 0.771 &0.763&0.736   \\
      $\Omega_m$&0.23-0.30&0.232&0.241&0.278 \\
      $\Omega_{b}$&0.04-0.05&0.040&0.041&0.050\\
      $h$&0.68-0.79&0.76&0.74&0.68\\
      $n_s$&0.9-0.99&0.97&0.97&0.96\\
      $\sigma_8$&0.7-0.83&0.73&0.74&0.73 \\
      & & & &\\
      \hline 
      $\chi^2$/$\nu$&-- &1.11&0.96&0.98\\
      \hline 
    \end{tabular}
  \end{center}
  \vspace{0.3cm}
  \label{tab_par}
\end{table}

The results of determination of the six main parameters of the
$\Lambda$CDM model are presented in Table~\ref{tab_par} for three
observational data sets: WMAP2006 data alone, data set
WMAP2006+BBN+h+SNIa, data set WMAP2006+BBN+h+SNIa+LSS. These models
are denoted as $\Lambda$CDM1, $\Lambda$CDM2 and $\Lambda$CDM3
correspondingly. For last we also computed 95.4\% confidence intervals
of the values of every parameter $p_{i} (i=1,2,...,6)$ by integrating
corresponding likelihood function
$L(p_{i})=exp[-0.5\Delta\chi^{2}(p_{i})]$ (see
\cite{novos2005b,apus2005} for details). As it was in paper
\cite{novos2005b} the likelihood functions of parameters are symmetric
with regard to the best-fit values and the Gaussian function is a good
approximation for it. Thus the values of the parameters listed in the
last column of Table~\ref{tab_par} lie in the middle of corresponding
ranges: $\Omega_{\Lambda} = 0.67$--$0.80$, $\Omega_{m} =
0.20$--$0.36$, $\Omega_{b} = 0.04$--$0.06$, $h = 0.59$--$0.76$, $n_{s}
= 0.945$--$0.975$, $\sigma_{8} = 0.65$--$0.81$. The values for the
$\Lambda$CDM1 and $\Lambda$CDM2 model parameters fit into this ranges
too. As one can see, the best-fit values of parameters obtained with
the use of the WMAP data alone coincide in practice with the results
of the WMAP team determination (Tables 2 and 5 in \cite{spergel2006}).
It is an important achievement for cosmology that the WMAP2006-based
determination of $h$, $\Omega_bh^2$ and $\Omega_m$ parameters agrees
with independent `direct' determinations by other authors
\cite{freedman2001,kirkman2003,per99}. This gives the ground the
$\Lambda$CDM model with parameters as estimated here or in
\cite{spergel2006} to be called the `concordance model' as it was
proposed by \cite{tegmark2000}. That is the reason why the inclusion
of direct measurements of parameters to the input data reduces the
value of $\chi^2/\nu$ relation.

The inclusion of data on the large-scale structure decreases the
best-fit value for the Hubble constant to the 0.68, but the values of
other parameters have remained within the limits of the standard
deviations $\sigma$ given by the WMAP team (Tables 2 and 5 in
\cite{spergel2006}). We should also point out on the `stability' of
the curvature parameter sign although its value is small and depends
on the data set used in parameters determination. It could indicate on
small but positive space curvature of observed Universe, so the
hyper-surface of constant time should be finite in volume and ever
expanding with increasing rate.

\section{WMAP2006: Large-scale structure}

\begin{figure}
  \includegraphics[width=13cm]{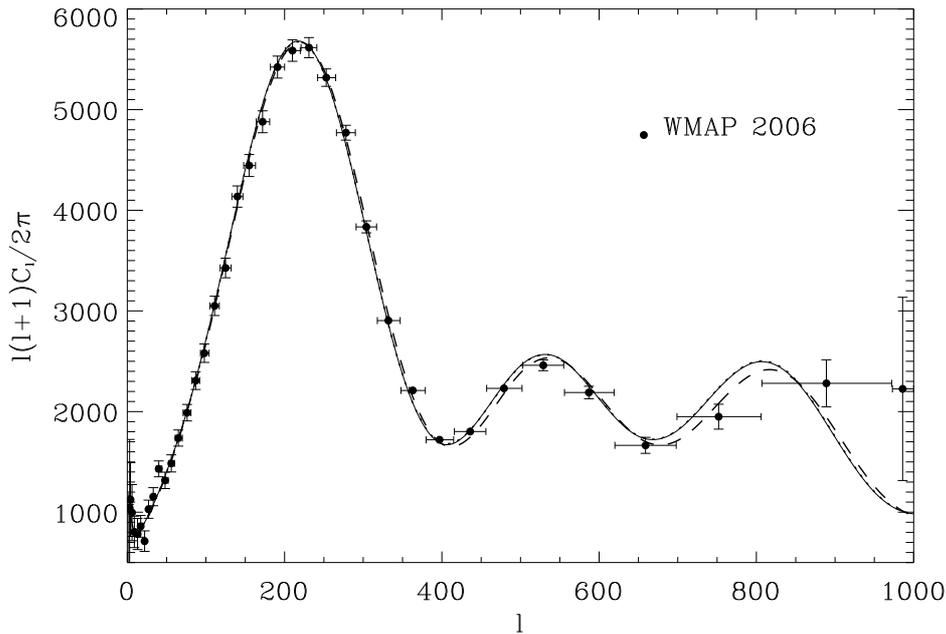}
  \caption{Power spectra of temperature fluctuations of CMB radiation
    in $\Lambda$CDM models with parameters as in Table~\ref{tab_par}
    (solid line for $\Lambda$CDM1, dash-dotted line for $\Lambda$CDM2,
    dashed for $\Lambda$CDM3), points denote the observed by WMAP2006
    spectrum.}
  \label{cmb}
\end{figure}

Let us compute with CMBfast code the angular power spectrum of CMB
temperature fluctuations, $\ell(\ell+1)C_\ell/2\pi$ for the
$\Lambda$CDM model with parameters listed in Table~\ref{tab_par} in
order to assess its accordance with observations. The spectrum is
normalized by minimization of $\chi^2$ with summation over
all points. The computation results are presented in Fig.~\ref{cmb},
$\chi^2=54.6$ for the spectrum of $\Lambda$CDM1 model, $\chi^2=53.3$
for $\Lambda$CDM2 and $\chi^2=37.2$ for $\Lambda$CDM3.  Thus, the
spectrum of $\Lambda$CDM3 model fits the observational points of the
spectrum with the same $\chi^2$ like in \cite{spergel2006} (in
Fig.~\ref{wmap3-sp} fits are overlayed).
The Table~\ref{tab_lss} lists the observed characteristics of
large-scale structure of the Universe, $\tilde y_j\pm \Delta \tilde
y_j$, they were used within procedure of determination parameters for
the $\Lambda$CDM model. Their values in $\Lambda$CDM model with
parameters which correspond to the minimum of the $\chi^2$ by different
determinations are quoted therein. The relative deviations of model
values from observational, $(y_j-\tilde y_j)/ \Delta \tilde y_j$, are
given in parenthesis. The positive sign means that the modelled
quantity is greater than observed one, negative sign means it is less.
An asterisk marks quantities which were not used in procedure of
parameters determination. As we can see, the most of characteristics
of large-scale structure of the Universe as evaluated on the basis of
$\Lambda$CDM model spectra fall within the 1$\sigma$ range around the
observational values and none of them fall outside of 3$\sigma$.
Statistics of deviations complies to the normal distribution.  Those
LSS characteristics, for which observational values deviate from
modelled for the 1$\sigma$ or more need more detailed analysis.

\begin{table}[ht]
  \caption{The characteristics of large-scale structure of Universe: 
    the observables along with their values for $\Lambda$CDM models 
    with parameters corresponding to the minimum of $\chi^2$ by 
    different determinations. The deviations are quoted in 
    parentheses for modelled quantities and observables in units of 
    1$\sigma$ error.}
  \label{tab}
  \begin{center}
    \begin{tabular}{|c|c|c|c|c|}
      \hline 
      Quantity&Observable&WMAP2006&WMAP2006+&WMAP2006+\\
      & & &BBN+h+SNIa&BBN+h+SNIa+LSS\\
      \hline
      & & & &\\
      $l_{p_1}$&$220.7\pm0.7$ \cite{hinshaw2006}&220.3  (-0.54)&220.4  (-0.43)&219.8  (-1.29) \\
      $(\Delta T)^2_{p_1}$, $(\mu K)^2$&$5619\pm30$ \cite{hinshaw2006}&5619  (0.00)&5615  (-0.13)&5604  (-0.51) \\
      $l_{d_1}$&$412.8\pm1.9$ \cite{hinshaw2006}&412.3  (-0.26)&412.0  (-0.42)&412.3  (-0.39) \\
      $(\Delta T)^2_{d_1}$, $(\mu K)^2$&$1704\pm27$ \cite{hinshaw2006}&1668  (-1.35)&1670  (-1.28)& 1673  (-1.15)\\
      $l_{p_2}$&$531.3\pm3.5$ \cite{hinshaw2006}&536.3  (+1.42)&536.2  (+1.39)&537.4  (+1.75) \\
      $(\Delta T)^2_{p_2}$, $(\mu K)^2$&$2476\pm40$ \cite{hinshaw2006}&2534  (+1.44)&2539  (+1.57)&2558  (+2.05)\\
      $l_{d_2}$&$674.6\pm12.1$ \cite{hinshaw2006}&674.4  (-0.02)&674.1  (-0.04)&679.7  (+0.42) \\
      $(\Delta T)^2_{d_2}$, $(\mu K)^2$&$1668\pm85$ \cite{hinshaw2006}&1692  (+0.28)&1700  (+0.38)&1687  (+0.22) \\
      $l_{p_3}$&$1143\pm167$ \cite{hinshaw2006}&$^{ *)}$814.5  (-1.97)&$^{ *)}$814.3  (-1.97)&$^{ *)}$820.2  (-1.93) \\
      $(\Delta T)^2_{p_3}$, $(\mu K)^2$&$2442\pm355$ \cite{hinshaw2006}&$^{ *)}$2451  (+0.03)&$^{ *)}$2463  (+0.06)&$^{ *)}$2439 (-0.01) \\
      $(\Delta T)^2_2$, $(\mu K)^2$& $211\pm 860$ \cite{hinshaw2006}&993  (+0.91)&980  (+0.89)&1013  (+0.93) \\
      $(\Delta T)^2_3$, $(\mu K)^2$& $1041\pm 664$ \cite{hinshaw2006}&940  (-0.15)&934  (-0.16)&983.4  (-0.93) \\
      $(\Delta T)^2_4$, $(\mu K)^2$& $731\pm 537$ \cite{hinshaw2006}&892  (+0.30)&889  (+0.29)&942.5  (+0.39) \\
      $(\Delta T)^2_5$, $(\mu K)^2$& $1521\pm 453$ \cite{hinshaw2006}&877  (-1.42)&873  (-1.43)&923.7  (-1.32) \\
      $(\Delta T)^2_6$, $(\mu K)^2$& $661\pm 395$ \cite{hinshaw2006}&851  (+0.48)&849  (+0.48)&903.4  (+0.61) \\
      $(\Delta T)^2_7$, $(\mu K)^2$& $ 1331\pm 353$ \cite{hinshaw2006}&835  (-1.41)&835  (-1.41)&892.5  (-1.24) \\
      $(\Delta T)^2_8$, $(\mu K)^2$& $671\pm 322$ \cite{hinshaw2006}&826  (+0.48)&827  (+0.48)&887.6  (+0.67) \\
      $(\Delta T)^2_9$, $(\mu K)^2$& $631\pm 298$ \cite{hinshaw2006}&822  (+0.64)&824  (+0.65)&886.6  (+0.86) \\
      $(\Delta T)^2_{10}$, $(\mu K)^2$& $751\pm 280$ \cite{hinshaw2006}&820  (+0.25)&843  (+0.26)&888.3  (+0.49) \\
      $h$&$0.72\pm0.08$ \cite{freedman2001}&$^{*)}$0.755  (+0.44)&0.744  (+0.30)&0.676  (-0.55)\\
      $\Omega_bh^2$&$0.0214\pm0.002$ \cite{kirkman2003}&$^{*)}$0.0228  (+0.70)&0.0228  (+0.71)&0.0228  (+0.72) \\
      $\Omega_m-0.75\Omega_{\Lambda}$&$-0.25\pm 0.125$ \cite{per99}&$^{*)}$-0.35  (-0.80)&-0.35  (-0.77)&-0.27  (-0.19) \\
      $V_{50}$, km/s&$370\pm 110$ \cite{dekel1999}&$^{*)}$251  (-1.08)&$^{*)}$255  (-1.05)&271  (-0.90) \\
      $\Delta_{\rho}$&$0.54\pm 0.13$ \cite{croft1999}&$^{*)}$0.57  (+0.20)&$^{ *)}$0.59  (+0.38)&0.55  (+0.10) \\
      $n_p$&$-2.47\pm 0.06$  \cite{croft1999}&$^{*)}$-2.49  (-0.37)&$^{ *)}$-2.49  (-0.35)&-2.50  (-0.55) \\
      $\Delta_{\rho}$&$0.72\pm0.09$\cite{mcdonald2000}&$^{*)}$0.57  (-1.67)&$^{ *)}$0.60  (-1.33)&$^{ *)}$0.56  (-1.79) \\
      $n_p$&$-2.55\pm 0.10$  \cite{mcdonald2000}&$^{*)}$-2.53  (+0.19)&$^{ *)}$-2.53  (+0.19)&$^{ *)}$-2.54  (+0.09) \\
      $\sigma_8$&$0.56\pm 0.071$ \cite{viana1999}&$^{*)}$0.36  (-2.82)&$^{*)}$0.38  (-2.54)&0.40  (-2.26) \\
      $\sigma_{cl}$&$0.508\pm 0.029$ \cite{pierpaoli2001}&$^{*)}$0.44  (-2.34)&$^{*)}$0.45  (-2.00)&0.46  (-1.33) \\
      \hline 
    \end{tabular}
  \end{center}
  $^{ *)}$ Not used in search procedure.
  \vspace{0.3cm}
  \label{tab_lss}
\end{table}

\begin{figure}
  \includegraphics[width=7.5cm]{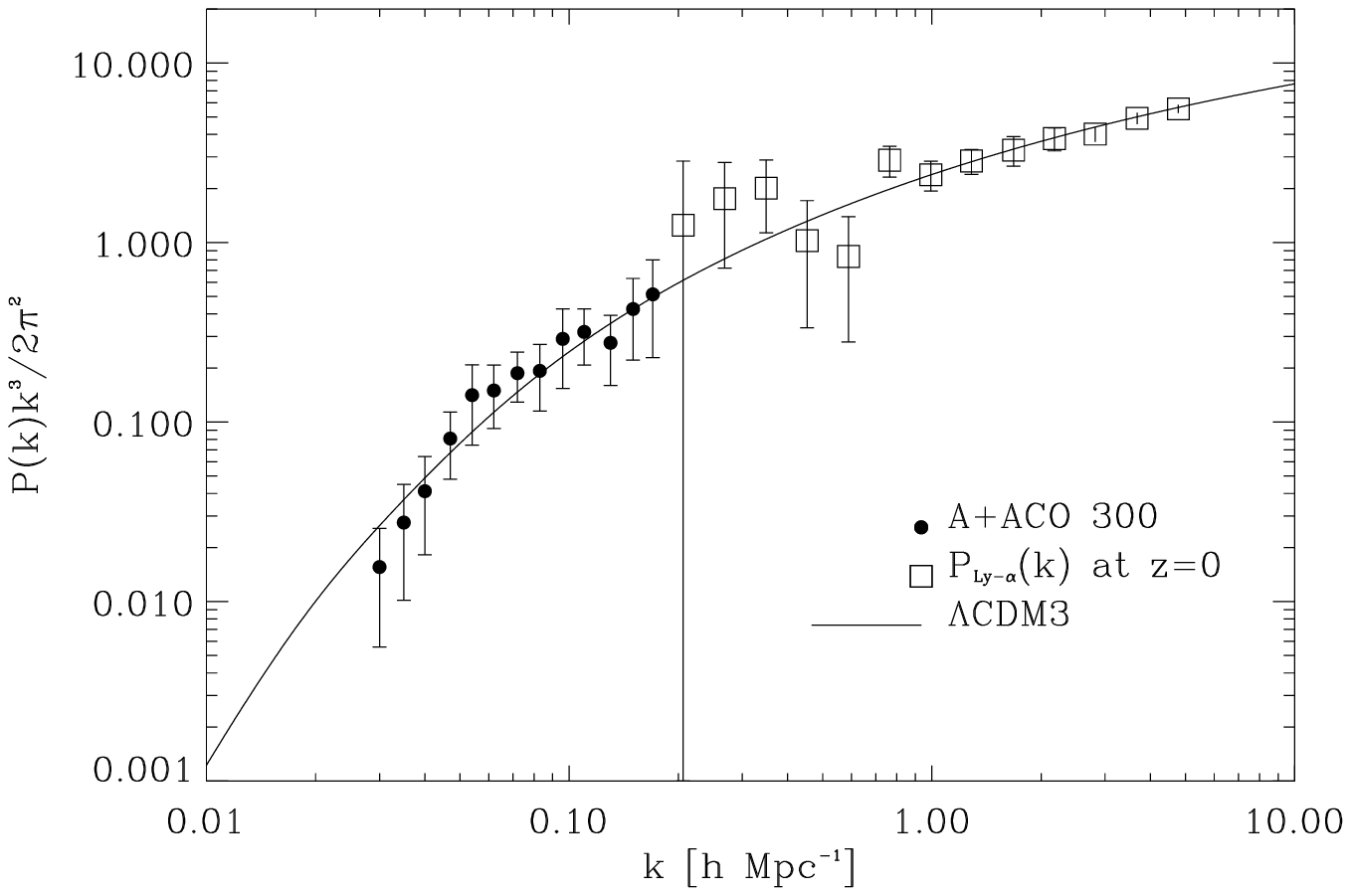}
  \includegraphics[width=7.5cm]{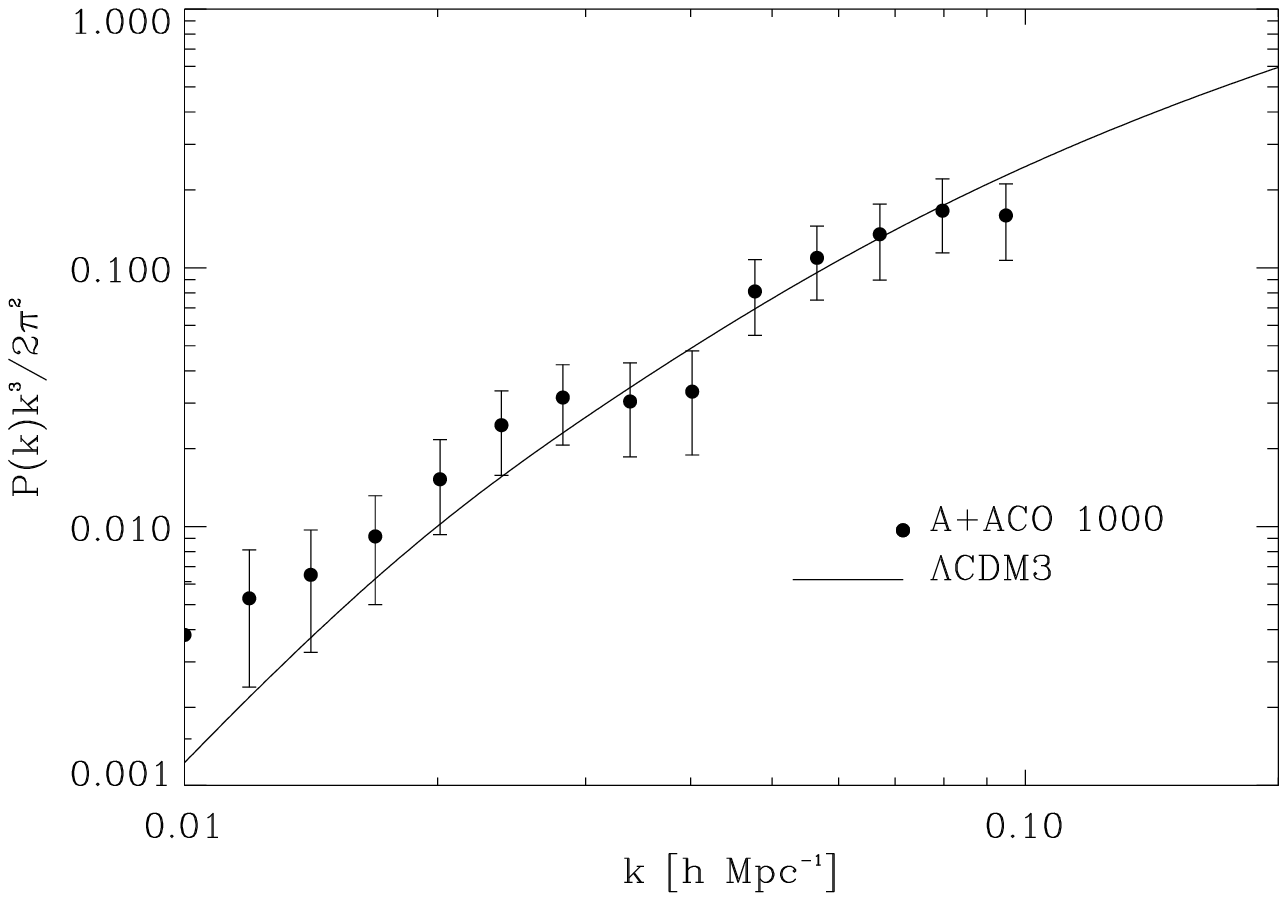}
  \caption{Power spectrum of matter density fluctuations as evaluated
    on the basis of spatial distribution of rich galaxy clusters from
    Abell-ACO catalog \cite{retzlaff1998}
    ($P(k)=P_{A+ACO}(k)/b^2_{A+ACO}$, $b_{A+ACO}=2.86$) as well as
    statistics of Ly$_\alpha$ absorption lines in the spectra of
    distant quasars \cite{croft1999} ($P_{Ly_{\alpha}}$). The power
    spectrum of matter fluctuations in $\Lambda$CDM3 model is plotted
    by solid line.}
  \label{cl_ly}
  \caption{Power spectrum of matter density fluctuations as evaluated
    on the basis of spatial distribution of rich galaxy clusters from
    Abell-ACO catalog \cite{miller2001}:
    $P(k)=P_{A+ACO}(k)/b^2_{A+ACO}$, $b_{A+ACO}=3.92$. The power
    spectrum of matter fluctuations in $\Lambda$CDM3 model is plotted
    by solid line.}
  \label{cl}
\end{figure}

Now let us compare the power spectrum of matter density perturbations
in the $\Lambda$CDM3 model with its estimates. The estimates were
obtained by analysis of inhomogeneities found in the spatial
distribution of galaxies and rich galaxy clusters. None of these
estimates was used in procedure of determination of parameters.  The
spectrum is presented as dimensionless quantity $\Delta^2(k)\equiv
P(k) k^3/2\pi^2$. It approximately equals to r.m.s. of matter density
perturbations averaged over region with size of $\pi/k$ Mpc. In a such
way we have built an independent criterion to estimate the validity of
parameters of the given model. In Fig.~\ref{cl_ly} two power spectra
are plotted, one is estimated from spatial correlation function of
rich galaxy clusters using the Abell and ACO catalogs
\cite{retzlaff1998} and the second from statistics of Ly$_\alpha$
absorption lines in the spectra of distant quasars, the lines are
originated by clouds of neutral hydrogen in intergalactic environment
\cite{croft1999}.  The experimental points of the second spectrum
virtually ``stack on'' the modelled spectrum in the part of small
scales ($k> 1h$). The oscillations seen within the range of scales
$0.2h\le k\le 0.8h$ come from of spatial limitations on sampling so
they do not represent the real behavior of the perturbation spectrum.
In the paper \cite{croft1999} the spectrum is given for $z=2.72$, so
we recalculated its amplitude to the $z=0$ using the evolution law for
perturbations from linear theory
$$
P(k)=P(k;z)D_1^2(0)/D_1^2(z),
$$
where $P(k;z)$ is the power spectrum for arbitrary $z$, $D_1(z)$ is a
linear factor of growth which is well approximated by an analytical
formula \cite{carrol1992}
$$
D_1(z)={5\over 2}{\Omega(z)\over 1+z}\left[{1\over
    70}+{209\Omega(z)-\Omega^2(z) \over
    140}+\Omega^{4/7}(z)\right]^{-1},
$$
$\Omega(z)=\Omega_m(1+z)^3/\left(\Omega_m(1+z)^3+\Omega_{\Lambda}\right)$.
Such concordance of the power spectrum of matter density perturbations
in the $\Lambda$CDM3 model with the same spectrum evaluated on the
basis of the observed data in \cite{croft1999} is a good reason for
model plausibility.

Since the bright galaxies and rich galaxy clusters were formed in
highest peaks of matter density perturbations the power spectrum of
inhomogeneous distribution of these objects relates to the spectrum of
matter density through the biasing parameter $b$:
$P(k)_{g,cl}=b^2_{g,cl}P(k)$ \cite{bardeen1986}. We determined $b$ for
the spectrum \cite{retzlaff1998} by minimization of deviations
$b^2\Delta^2(k)$ from $\Delta^2_{A+ACO}(k_I)$ using the
Levenberg-Markquardt method: $b_{A+ACO}=2.86$.  As one can see from
Fig.~\ref{cl_ly} the shape of spectrum $P_{A+ACO}(k)/b^2_{A+ACO}$ is
well approximated by the spectrum of the $\Lambda$CDM3 model. With
increase of scale ($k$ decreases) in the range $k\le 0.05h$
$\Delta^2_{A+ACO}$ falls down steeper than $\Delta^2_{\Lambda CDM}$.
The cause is that the sample of rich galaxy clusters used in
\cite{retzlaff1998} is spatially limited ($\le 300h^{-1}$Mpc), a
similar spectrum from sample with 1 Gpc scale obtained in
\cite{miller2001} and presented in Fig~\ref{cl} confirms that. The
scale-independent biasing parameter $b_{cl}=3.92$ therein was
determined in the same way as in previous case. The apparent rise of
the spectrum on large scales most probably is caused by dependence of
bias upon a scale, that was not taken into account. Indeed, the
brighter clusters get an advantage on large distances, they was formed
in higher peaks of matter density perturbations so they look more
clustered (see for example \cite{tegmark2004}).

\begin{figure}[htb]
\begin{center}
  \includegraphics[width=7.5cm]{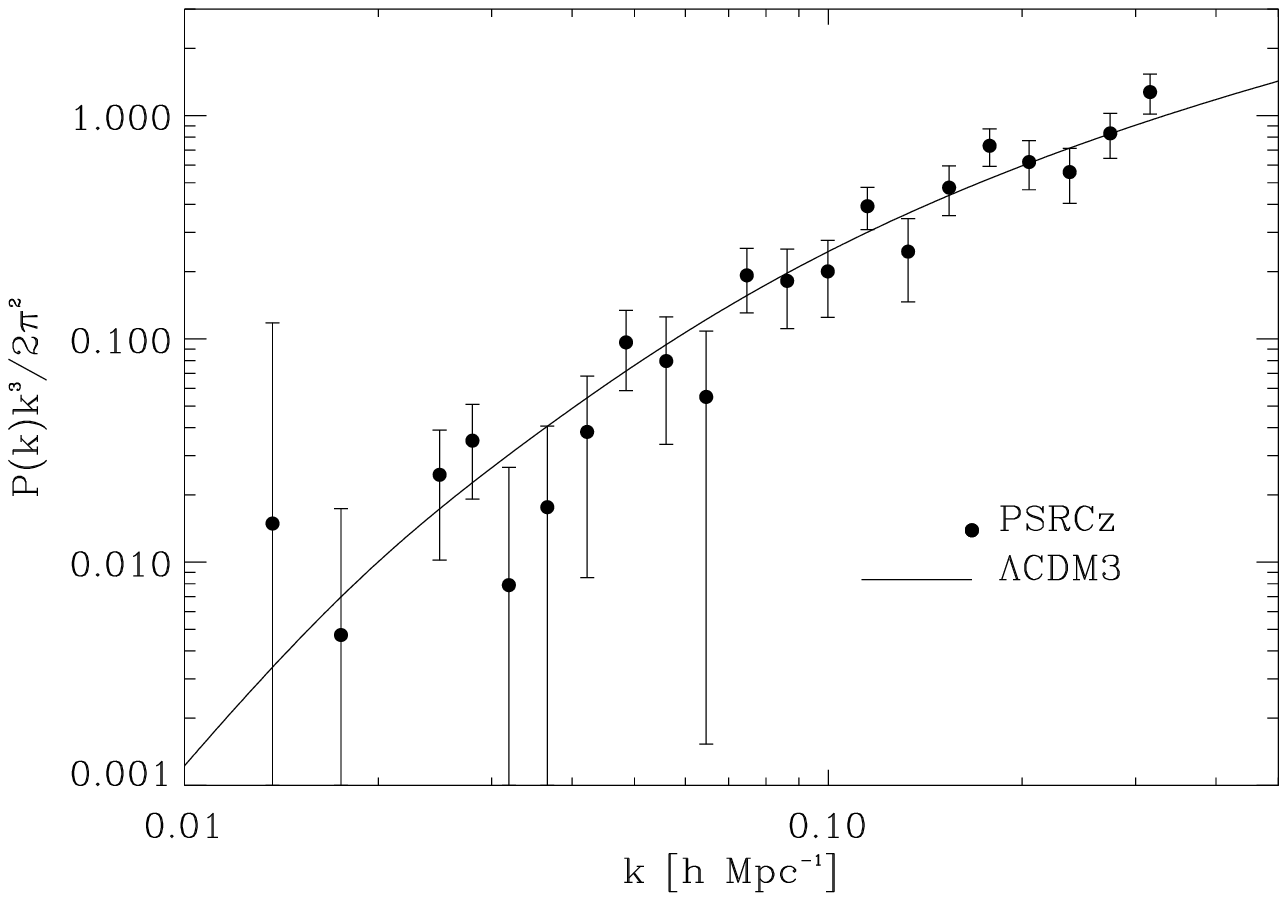}
  \includegraphics[width=7.5cm]{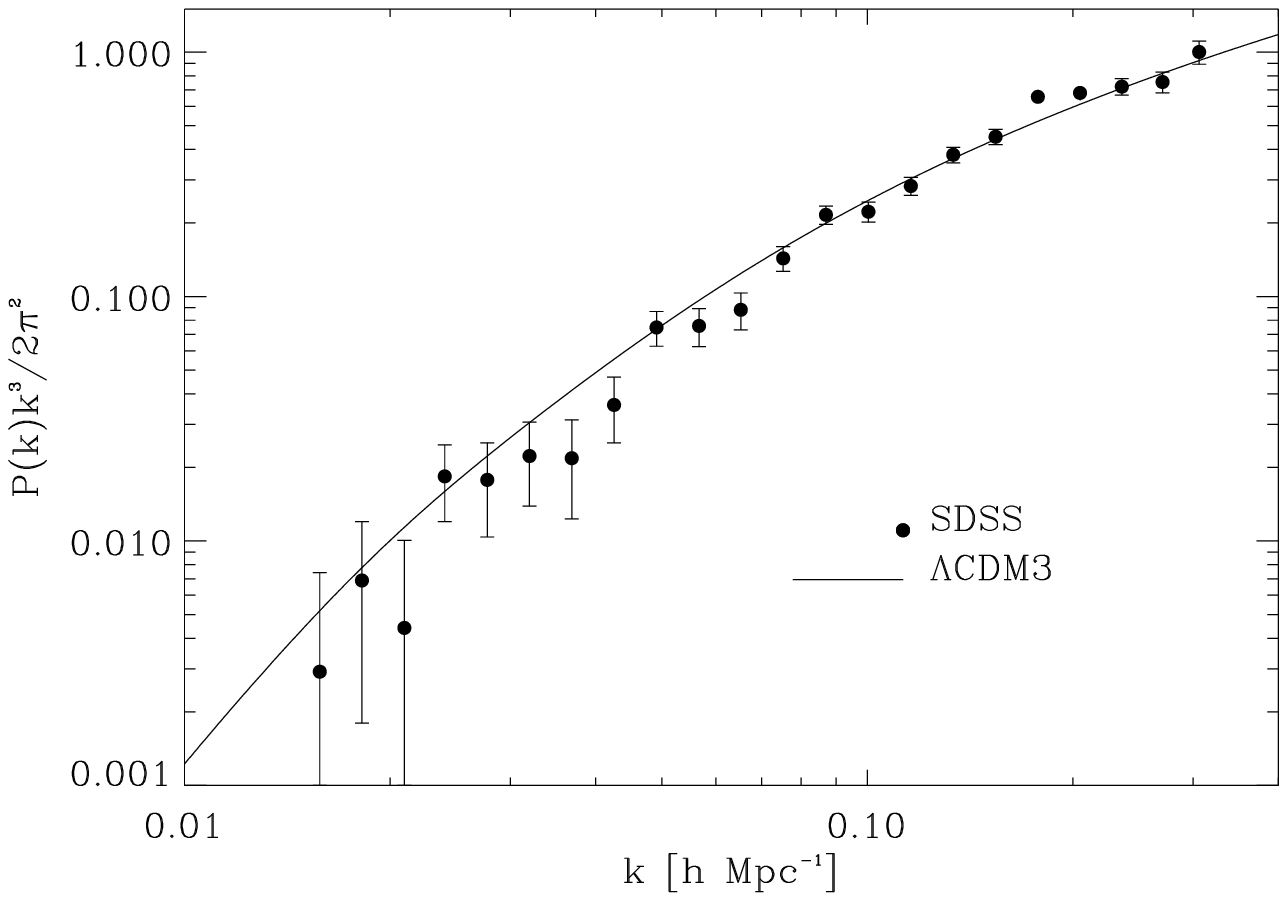}\\
  \caption{The power spectrum of matter density fluctuations as
    obtained on the basis of spatial distribution of {\it IRAS} galaxies
    in Point Source Catalog Redshift Survey (PSCRz)
    \cite{hamilton2002}.  $P(k)=P_{PSRCz}(k)/b^2_{PSRCz}$,
    $b_{PSRCz}=1.08$. The power spectrum of density fluctuations in
    $\Lambda$CDM3 model is plotted by solid line.}
  \label{pscz}
  \caption{The power spectrum of matter density fluctuations as
    obtained on the basis of spatial distribution of galaxies in Sloan
    Digital Sky Survey (SDSS) \cite{tegmark2004}.
    $P(k)=P_{SDSS}(k)/b^2_{SDSS}$, $b_{SDSS}=1.21$. The power spectrum
    of density fluctuations in $\Lambda$CDM3 model is plotted by solid
    line.}
  \label{sdss}
\end{center}
\end{figure}

At the next step we compare the model spectra with the spectra
obtained on the basis of observed spatial distribution of galaxies.
The treatment of galaxies power spectra at small scales ($k>0.3h$
Mpc$^-1$) is complicated by nonlinear distortions, the
scale-dependence of biasing and by correlation between the values of
its amplitude in the adjacent wave-number ranges as well. Therefore we
have restricted ourself to the section of galactic spectrum starting
from $k\le 0.3h$ Mpc$^{-1}$ when doing comparison with linear spectrum
of the $\Lambda$CDM model. Also we assume biasing to be independent on
$k$ in this range of scales. The special techniques were proposed in
\cite{hamilton2002} for decorrelating data points. It allows to
decrease the correlations in the range of nonlinearity and virtually
removes them in the linear range.  The decorrelated power spectrum of
the PSRCz survey made of IRAS galaxies (Point Source Redshift Catalog)
is presented (see Fig.~\ref{pscz}).  We have determined the biasing
parameter for this spectrum by $\chi^2$-minimization by the
Levenberg-Markquardt method and it has appeared to be close to unity,
$b_{PSRCz}=1.08$, the proper value for galaxies from IRAS survey.  The
same power spectrum for Sloan Digital Sky Survey (SDSS) is presented
in Fig.~\ref{sdss}. The biasing parameter is $b_{SDSS}=1.21$.  The
amplitudes of the galactic spectra were divided by squared
corresponding biasing factors to make the comparison with linear power
spectrum of $\Lambda$CDM model. The linear power spectrum of matter
density perturbations of the $\Lambda$CDM model conforms quite well to
the amplitudes and shapes (the dependence on $k$) of the observed
power spectra of inhomogeneities in the spatial distributions of
galaxies.

\section{Discussion of results and conclusions}

Now we assess the concordance of two theoretical spectra with observed
ones for two models the above mentioned $\Lambda$CDM3 and $\Lambda$CDM
model from \cite{spergel2006} with parameters $\Omega_{\Lambda}=0.76$,
$\Omega_m=0.24$, $\Omega_{b}=0.042$, $h=0.73$, $A_s=0.83$,
$n_s=0.958$.  The values of $\chi^2$ for all these power spectra are
listed in Table~\ref{tab_chi}. The values of biasing factors computed
in the same way for both models are presented in parentheses for the
power spectra from spatial distribution of galaxies and clusters. As
it can be seen, the power spectrum of $\Lambda$CDM3 model provides the
fit to the most of observed spectra with smaller value of $\chi^2$
(except for $P_{A+ACO}(k)$ \cite{miller2001}) than $\Lambda$CDM model
does \cite{spergel2006}. Thus, among two models the $\Lambda$CDM3
model could be deemed as closer to true model of Universe, which is
still searched.

\begin{table}[ht]
  \caption{$\chi^2$ for different spectra of $\Lambda$CDM3 and $\Lambda$CDM models \cite{spergel2006}.}
  \begin{center}
    \begin{tabular}{|c|c|c|}
      \hline 
      & $\chi^2$ × $\Lambda$CDM\cite{spergel2006} & $\chi^2$  × $\Lambda$CDM3 \\
      \hline 
      $\ell(\ell+1)C_\ell/2\pi$ \cite{hinshaw2006}&37.8&37.2\\
      $P_{Ly_{\alpha}}(k)$ \cite{croft1999}&17.81 & 8.13\\
      $P_{A+ACO}(k)$ \cite{retzlaff1998}&4.22 ($b=2.70$)&4.02 ($b=2.86$)\\
      $P_{A+ACO}(k)$ \cite{miller2001}&8.44 ($b=3.70$)&9.43 ($b=3.92$)\\
      $P_{PSRCz}(k)$ \cite{hamilton2002}&14.68 ($b=1.03$)&14.03 ($b=1.08$)\\
      $P_{SDSS}(k)$ \cite{tegmark2004}&46.93 ($b=1.16$)&39.29 ($b=1.21$)\\
      \hline 
    \end{tabular}
  \end{center}
  \vspace{0.3cm}
  \label{tab_chi}
\end{table}

Basing on confrontation of predicted for $\Lambda$CDM model
characteristics of LSS with observed ones as presented in
Table~\ref{tab_lss} and Fig.~\ref{cmb}--\ref{sdss} we should point out
that regardless of pretty good concordance of the data set in a whole
there are still ``stable'' deviations of some quantities beyond the
$1\sigma$ confidence limits.  The deviation is called ``stable'' if it
can not be removed by the change of set of the observational data.
Namely, they are:
\begin{itemize}
\item the positions of 2nd and 3rd acoustic peaks, $\ell_{p_2}$,
  $\ell_{p_3}$;
\item amplitudes of 1st dip, 2nd and 3rd acoustic peaks, $(\Delta
  T)^2_{d_1}$, $(\Delta T)^2_{p_2}$, $(\Delta T)^2_{p_3}$;
\item amplitudes of 5th and 7th spherical harmonics, $(\Delta T)^2_5$,
  $(\Delta T)^2_7$;
\item amplitude of the power spectrum of density perturbations based
  of X-ray temperature function ($\sigma_8$ \cite{viana1999}) and mass
  function of rich galaxy clusters ($\sigma_{cl}$
  \cite{pierpaoli2001}).
\end{itemize}

The following explanations could be proposed for these deviations: i)
excessive ``stiffness'' of $\Lambda$CDM model, ii) an assumption of
scale-invariance of primordial perturbations spectrum, iii) the
underestimation of $1\sigma$ C.L. for some of experimental quantities.
Obviously, the extension of the $\Lambda$CDM models towards inclusion
of dark energy or inflation models giving scale-dependent power
spectrum demands for to the possibility of verification by
high-quality observational data.

Hence, the data of observational cosmology spreading over scales from
$1$\,Mpc to $10000$\,Mpc indicate that $\Lambda$CDM model with
parameters $\Omega_{\Lambda}=0.736$, $\Omega_m=0.278$,
$\Omega_{b}=0.05$, $h=0.68$, $\sigma_8=0.73$ and $n_s=0.96$ is the
best-fit for whole data set. So this model can be considered as the
closest to the true model of Universe within the class of 6-parameter
cosmological models.

\end{document}